\documentclass[pra,twocolumn,showpacs,preprintnumbers,amsmath,amssymb]{revtex4}
\usepackage{mathrsfs}
\usepackage{bbm}
\usepackage{amsfonts}
\usepackage{tipa}

\usepackage{epsfig,graphicx}
\usepackage{amstext}
\usepackage{amsmath}
\usepackage{graphicx}
\usepackage{times}
\usepackage{graphicx}

\begin{document}


\title{Measurement-induced nonlocality in the anisotropic Heisenberg chain}
\author{Wen-Xue Chen}
\author{Yu-Xia Xie}
\author{Xiao-Qiang Xi}
\email{xxq@xupt.edu.cn}
\address{School of Science, Xi'an University of Posts and
         Telecommunications, Xi'an 710121, China}

\begin{abstract}
 Quantum correlations are essential for quantum information processing.
 Measurement-induced nonlocality (MIN) which is defined based on the
 projective measurement is a good measure of quantum correlation, and is
 favored for its potential applications. We investigate here behaviors of
 the geometric and entropic MIN in the two-qubit Heisenberg XY chain, and
 reveal effects of the anisotropic parameter $\gamma$ as well as
 the external magnetic field $B$ on strength of them. Our results
 show that both $\gamma$ and $B$ can serve as efficient controlling
 parameters for tuning the MIN in the XY chain.
\end{abstract}


\maketitle

\section{Introduction}\label{sec:1}
Quantum correlations plays an important role for quantum protocols
which outperform those of the classical one. The typical protocols
are those proposed in the literature, such as the quantum
cryptography \cite{Ekert}, teleportation
\cite{teleport1,teleport2,teleport3}, and quantum computation
\cite{Bennett}, which depend crucially on the entanglement between
the considered system. While quantum entanglement has been
recognized as crucial for various quantum information processing
tasks, recent investigations show that it is in fact a special kind
of quantum correlation measure, and there are other forms of quantum
correlations which are more fundamental than that of entanglement
\cite{RMP}. The typical and seminal one are the concept of quantum
discord proposed by Ollivier and Zurek \cite{Ollivier}, and
Henderson and Vedral \cite{Henderson}, which can exist even for the
separable states, and is considered to be responsible for the power
of the deterministic quantum computation with one qubit
\cite{Knill,Datta,Lanyon}. Due to these reasons and its fundamentals
in quantum mechanics \cite{Huml,Gumile,Hu2,Hu3}, great efforts has
been devoted to this attractive field in the pat few years.

The quantum discord, although is favored for its operational
interpretation and possible application, is very hard to calculate
\cite{Girolami}. To avoid this problem, Luo and Fu
\cite{Luof1,Luof2} introduced a geometric description of quantum
correlation which they termed as measurement-induced nonlocality
(MIN). As its name implies, the MIN can also be considered as a
measure of nonlocality for it is defined via the maximal global
disturbance of the locally invariant measurements, and it is in some
sense more general than that of the Bell-type nonlocality. Moreover,
the geometric MIN can also be evaluated analytically for the
two-qubit states.

Besides the geometric MIN, an entropic measure of MIN based on the
von Neumann entropy was introduced recently \cite{min1}, which is
equivalent to that defined via the relative entropy \cite{min2}.
Although its calculation is harder than that of the geometric one,
it was favored for its definite physical meaning, which can be
interpreted as the maximal increment of one's uncertainty about a
system after his local invariant measurements on one subsystem.

In this work, we study effects of the anisotropic parameter and the
external magnetic field on the above two scenarios of MIN for a
Heisenberg XY chain. The structure of this paper is arranged as
follows. In Section \ref{sec:2}, we recall the concept of the
geometric and entropc MIN, and the model of the XY chain we
considered. Then in Section \ref{sec:3}, we examine behaviors of the
MIN in the XY chain with different system parameters. We summarize
the main finding of this work in Section \ref{sec:4}.

\section{Basic formalism for MIN and the model}\label{sec:2}
Here we first recall the concept of MIN. In its original geometric
quantification scheme \cite{Luof2}, it was defined by the square of
the maximum of the Schatten 2-norm between two states $\rho$ and
$\Pi^A(\rho)$, which corresponds to the bipartite states before and
after the projective measurements. To be explicitly, we can write it
as follows
\begin{eqnarray}
 N^s(\rho)=\max_{\Pi^A}||\rho-\Pi^A(\rho)||^2,
\end{eqnarray}
with the maximum being taken over the local projective measurements
$\Pi^a=\{\Pi_k^a\}$ on subsystem $A$, which satisfy the condition
$\sum_k \Pi_k^A\rho^A\Pi_k^A=\rho^A$. Moreover, $||X||$ is the
Schatten 2-norm where $||X||^2=\text{Tr}(X^\dag X)$.

For the $2\times n$ dimensional states, solutions of the geometric
MIN can be derived analytically. Here for convenience of
representation, we do not list its explicit expressions, and the
readers who are interested in it can resort to the Ref. \cite{Luof2}
for more detail.

Different from $N^s(\rho)$ which measures the MIN from a geometric
perspective, the entropic measure of MIN is defined as follows
\cite{min1}
\begin{eqnarray}
 N^v(\rho)=I(\rho)-\min_{\Pi^a}I[\Pi^a(\rho)],
\end{eqnarray}
where $I(\rho)=S(\rho^A)+S(\rho^B)-S(\rho)$ denotes the quantum
mutual information, and the minimum is taken over the same
projective measurements as that in the definition of $N^s(\rho)$.
This MIN quantifies the maximal loss of total correlations under
non-disturbing local measurements, and it is equivalent to
$\max_{\Pi^A}S[\Pi^a(\rho)]-S(\rho)$ because of the definition of
$\Pi^A$ which leaves $\rho^A$ invariant.

The evaluation of $N^v(\rho)$ is also difficult due to the
optimization processor involved. But for the special case that the
bipartite states $\rho$ having nondegenerate $\rho^A$, the optimal
$\Pi^A$ for obtaining the entropic MIN correspond to that of the
spectral resolutions of $\rho^A=\sum_k p_k\Pi_k^A$, and thus
$N^v(\rho)$ can be evaluated easily.

After review of the basic formalism for MIN, we present the model we
considered in this paper. This is the usual two-spin Heisenberg XY
chain with uniform external magnetic field applies along the $z$
direction. The corresponding Hamiltonian is given by \cite{chain}
\begin{eqnarray}
 \hat{H}&=&J(S_1^+S_2^-+S_1^-S_2^+)+J\gamma(S_1^+S_2^++S_1^-S_2^-)\nonumber\\
         &&+B(S_1^z+S_2^z),
\end{eqnarray}
where $B$ denotes the strength of the external magnetic field, and
$\gamma$ measures the anisotropy of the system which reduces to the
XX model for $\gamma=0$, and the Ising model for $\gamma=1$.

After a straightforward algebra, the eigenvalues and eigenvectors of
$\hat{H}$ can be derived analytically as
\begin{eqnarray}
 \varepsilon_{1,2}=\pm J,~\varepsilon_{3,4}=\pm \eta,
\end{eqnarray}
and
\begin{eqnarray}
 &&|\psi\rangle_{1,2}=\frac{1}{\sqrt{2}}(|01\rangle\pm
 |10\rangle),\nonumber\\
 &&|\psi\rangle_{3,4}=u^\pm(|00\rangle \pm \frac{J\gamma}{\eta \mp
 B}|11\rangle),
\end{eqnarray}
where $\eta=\sqrt{B^2+J^2\gamma^2}$, and the normalization parameter
$u^\pm=(\eta\mp B)/\sqrt{J^2\gamma^2+(\eta\mp B)^2}$.

Based on the above eigenvalues and eigenvectors, we can now evaluate
the MIN in the XY chain, and for this purpose, we need to introduce
the notion of the thermal state $\rho=Z^{-1}\exp{(-\hat{H}/{k_B
T})}$, with $Z=\text{Tr}[\exp{(-\hat{H}/{k_B T})}]$ being the
partition function, $k_B$ is the Boltzmann constant, and $T$ denotes
the temperature. When $T=0$, $\rho$ reduces to the so-called ground
state.

\section{MIN in the Heisenberg XY chain}\label{sec:3}
\begin{figure}
\centering
\resizebox{0.4\textwidth}{!}{%
\includegraphics{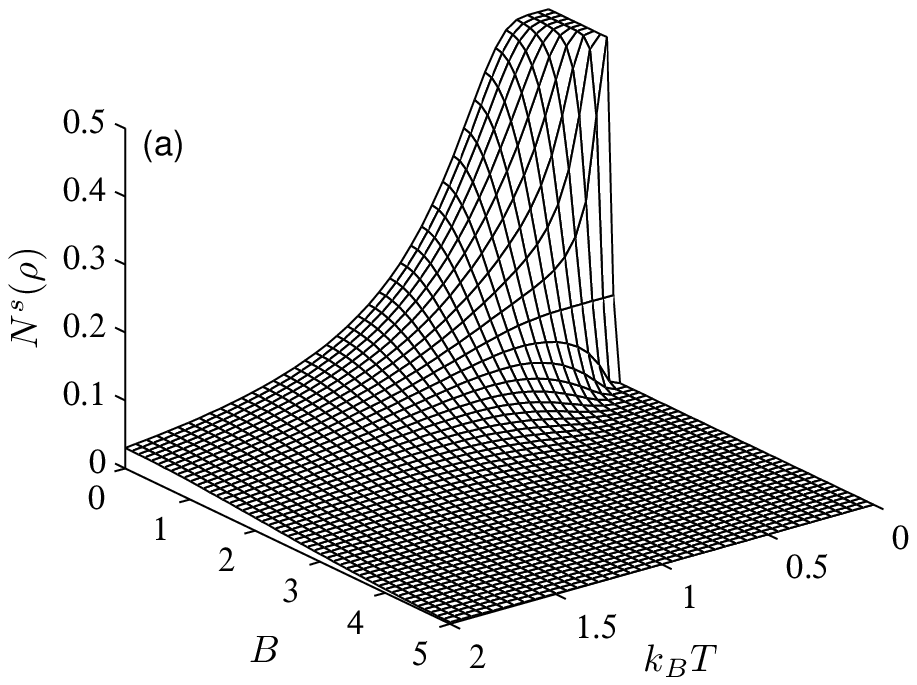}}
\centering
\resizebox{0.4\textwidth}{!}{%
\includegraphics{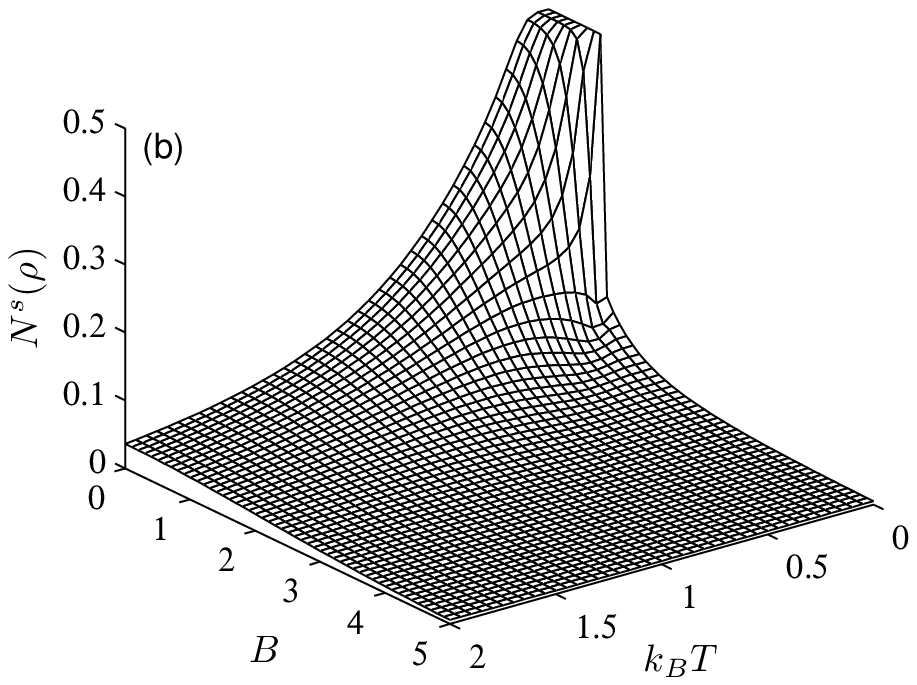}}
\centering
\resizebox{0.4\textwidth}{!}{%
\includegraphics{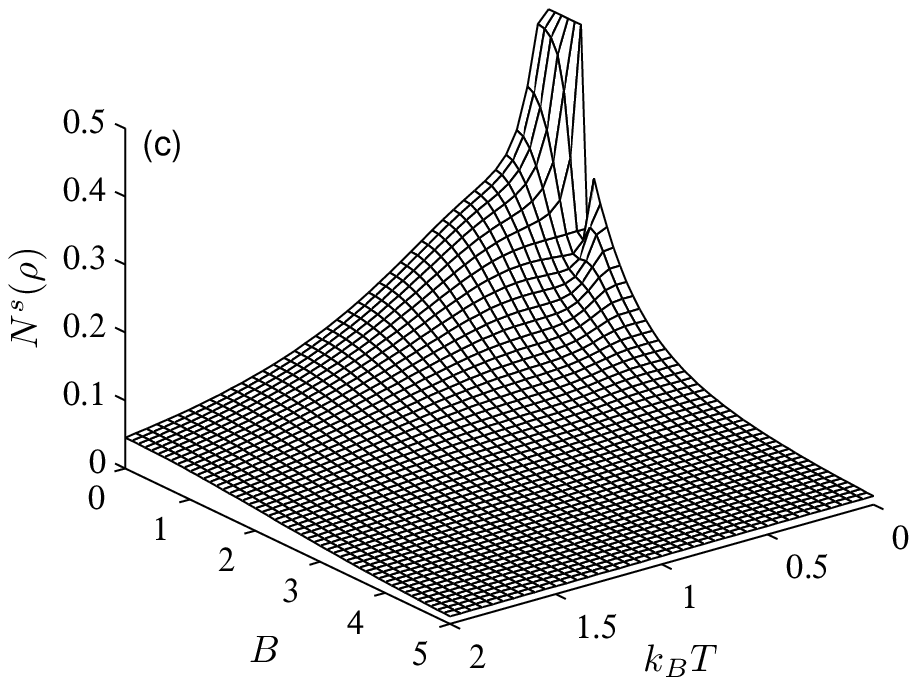}}
\centering
\resizebox{0.4\textwidth}{!}{%
\includegraphics{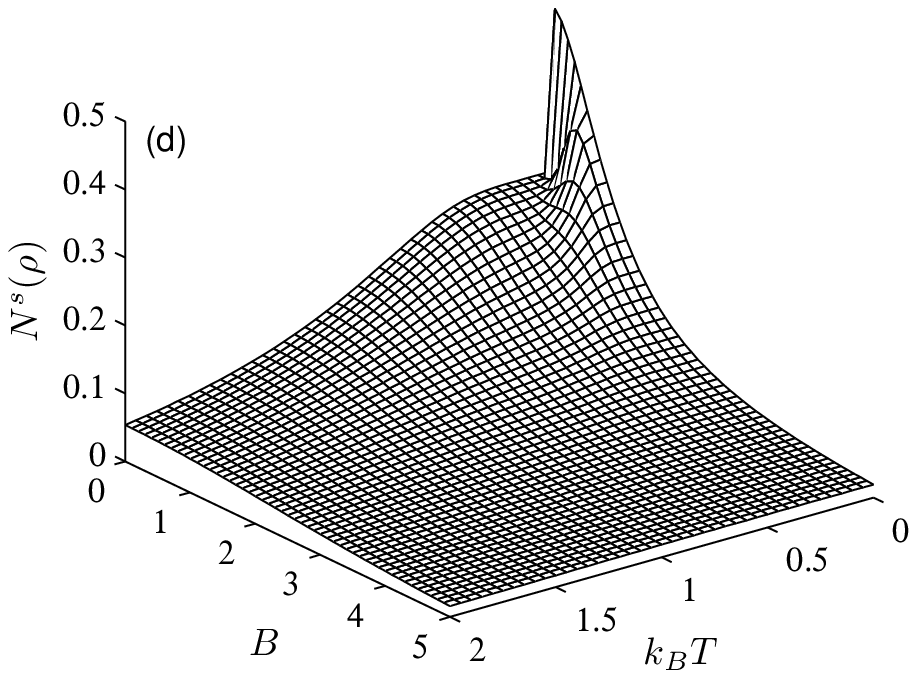}}
\caption{$N^s(\rho)$ versus $B$ and $k_B T$ for the Heisenberg XY
         chain with the anisotropic parameters $\gamma=0$ (a),
         $\gamma=0.5$ (b), and $\gamma=0.8$ (c), and $\gamma=1$ (d),
         respectively. Moreover, the parameter $J$ in these plots
         is chosen to be unity.}
        \label{fig:1}
\end{figure}

\begin{figure*}[htbp]
\centering
\resizebox{0.4\textwidth}{!}{%
\includegraphics{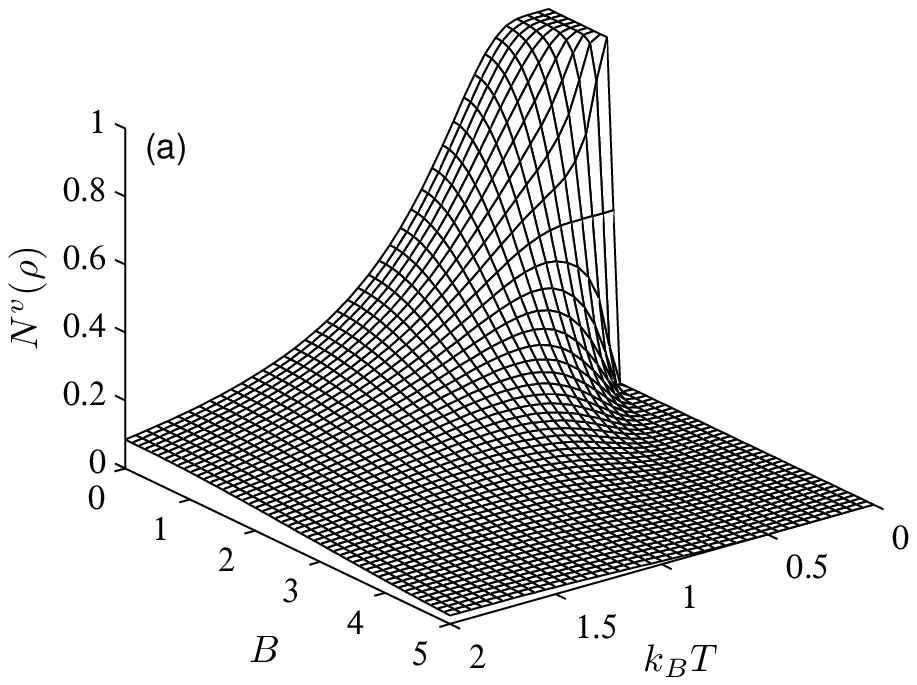}}
\centering
\resizebox{0.4\textwidth}{!}{%
\includegraphics{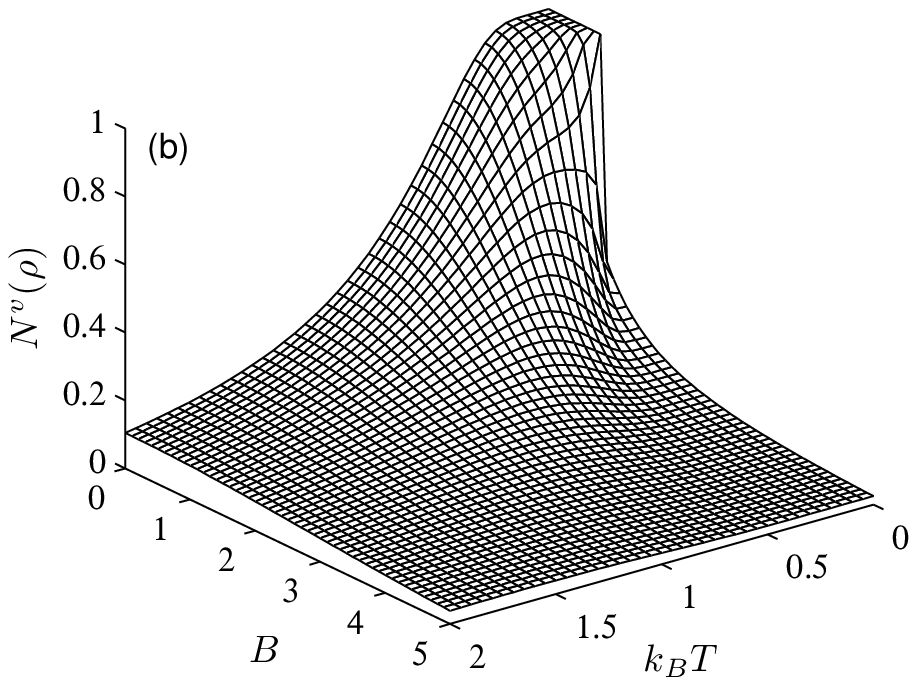}}
\centering
\resizebox{0.4\textwidth}{!}{%
\includegraphics{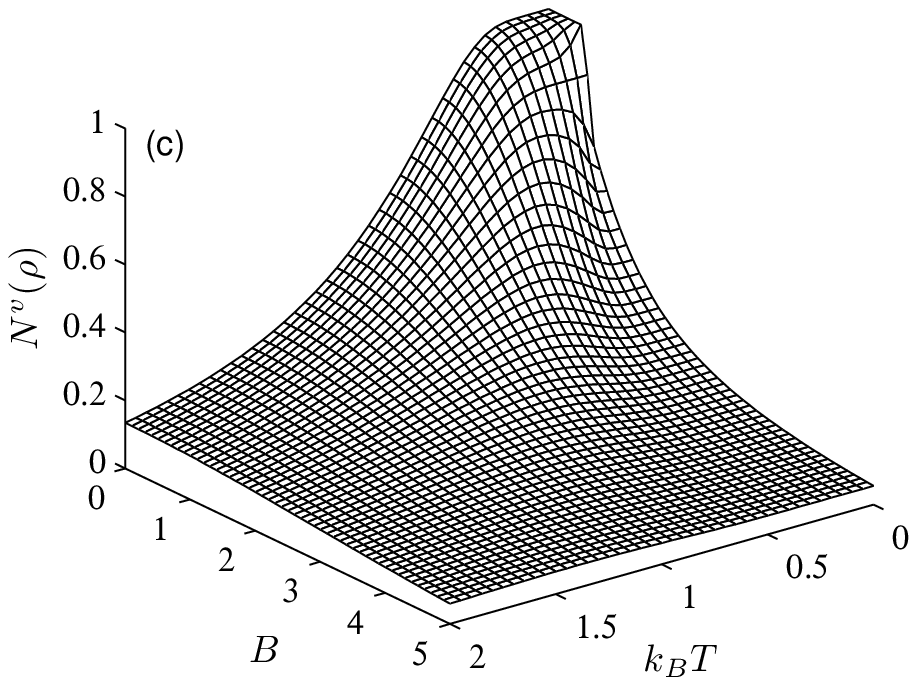}}
\centering
\resizebox{0.4\textwidth}{!}{%
\includegraphics{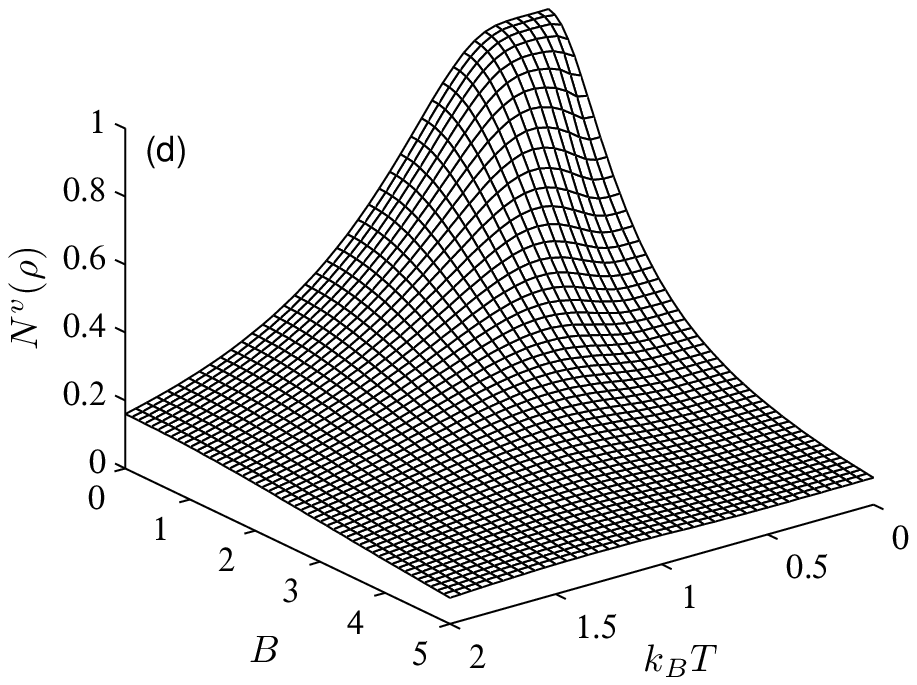}}
\caption{$N^v(\rho)$ versus $B$ and $k_B T$ for the Heisenberg XY
         chain with the anisotropic parameters $\gamma=0$ (a),
         $\gamma=0.5$ (b), and $\gamma=0.8$ (c), and $\gamma=1$ (d),
         respectively. Moreover, the parameter $J$ in these plots
         is chosen to be unity.}
        \label{fig:2}
\end{figure*}

We now begin our discussion about behaviors of MIN in the XY chain.
To be explicitly, we will evaluate influence of the anisotropic
parameter $\gamma$, strength of the external magnetic field $B$, and
the temperature $T$ on MIN. We first consider the scenario of the
geometric MIN. In Fig. \ref{fig:1}, we gave three-dimensional plots
of the geometric MIN $N^s(\rho)$ as functions of $B$ and $k_B T$ for
various values of $\gamma$. When considering the case of ground
states at absolute zero temperature, one can see clearly that for
the isotropic XY chain (i.e., $\gamma=0$, for which it is also
termed as the XX chain), $N^s(\rho)$ keeps the constant value of 0.5
before the critical point $B_c=J\sqrt{1-\gamma^2}$, and disappears
suddenly when $B>B_c$. This sudden death behavior occurs only for
the case of $\gamma=0$, and for finite $\gamma$ as displayed in Fig.
\ref{fig:1}(b), $N^s(\rho)$ maintains the constant value 0.5 before
$B_c$, and decreases exponentially to zero in the infinite limit of
$B$. But when the anisotropic parameter $\gamma$ exceeds a critical
point $\gamma_c$, such as that displayed in Fig. \ref{fig:1}(c) with
$\gamma=0.8$, $N^s(\rho)$ initially keeps the maximum value 0.5, and
then decreases suddenly to a finite value $N^s(\rho)<0.5$ at
$B=B_c$, after which it revivals and then decreases gradually to
zero. This is one of the main difference between $N^s(\rho)$ in the
regions of $\gamma<\gamma_c$ and $\gamma>\gamma_c$, which implies
that the anisotropic parameter $\gamma$ can serve as an efficient
parameter for tuning magnitudes of the MIN. Finally, we displayed in
Fig. \ref{fig:1}(d) behaviors of $N^s(\rho)$ for the special case of
$\gamma=1$ which corresponds to the Ising chain. Different from
those for $\gamma<1$, here $N^s(\rho)$ decreases gradually from its
maximum 1 to zero with increasing magnetic field $B$.

For thermal states with finite temperatures, one may expects that
the MIN $N^s(\rho)$ may be decreased by increasing the temperature,
and this is indeed the case for weak magnetic fields. But there is a
peculiar phenomenon needs to be pay attention to. As can be observed
from Fig. \ref{fig:1}(a) and (b), the MIN may be increased by
increasing temperature in the region of relative strong external
magnetic field. This shows again the combined effects of temperature
and the anisotropic parameter on the geometric MIN.

From the above discussion, one can see that the geometric MIN
$N^s(\rho)$ is sensitive to the parameter $\gamma$ and $B$. We now
turn to investigate behaviors of the entropic MIN in the XY chain.
The exemplified plots are displayed in Fig. \ref{fig:2}, where the
entropic MIN $N^v(\rho)$ are plotted as functions of magnetic field
$B$ and the scaled temperature $k_B T$ with different anisotropic
parameters $\gamma$. From these plots one can observe that for
relative small values of $\gamma$, e.g., the case of $\gamma=0$ as
shown in Fig. \ref{fig:2}(a) and $\gamma=0.5$ as shown in Fig.
\ref{fig:2}(b), $N^v(\rho)$ exhibit similar behaviors as those for
$N^s(\rho)$ with the same parameters [see, Fig. \ref{fig:1}(a) and
(b)], and the only difference is that the constant value for
$N^v(\rho)$ is twice that of $N^s(\rho)$. For large values of
$\gamma$, however, $N^v(\rho)$ and $N^s(\rho)$ exhibit completely
different behaviors, which is more evident in the low temperature
region. As can be seen from Fig. \ref{fig:2}(c) and (d), $N^v(\rho)$
is an non-increasing function of $B$ for any fixed $k_B T$, and
there are no revival of $N^v(\rho)$, even at zero absolute
temperature. This is the first difference between geometric and
entropic description of MIN in the Heisenberg XY chain. Besides
this, one can also note that the MIN $N^v(\rho)$ can be increased by
increasing temperature in the whole region of $\gamma$ with
appropriate chosen magnetic fields, and this constitutes another
main difference between $N^v(\rho)$ and $N^s(\rho)$ in describing
the nonlocal property of the Heisenberg XY chain.

Before ending this section, we would like to point out that here the
different behaviors between $N^v(\rho)$ and $N^s(\rho)$ for the
Heisenberg XY chain is caused by the different metric adopted for
defining the MIN. This reveals the relativity of different quantum
correlation measures in quantifying a specific quantum system
\cite{Hu2}, and the searching of an universal scheme for quantifying
correlations remains an open problem needs to be solved in future.

\section{Summary and discussion}\label{sec:4}
To summarize, we have studied influence of the anisotropic parameter
$\gamma$ and strength of the external magnetic field $B$ on
behaviors of the MIN. Through detailed calculation and comparison,
we showed that both of them can serve as efficient parameters for
controlling magnitudes of MIN in the Heisenberg XY chain. We also
showed that the geometric measure of MIN and the entropic measure of
MIN may correspond to completely different behaviors of nonlocal
correlations, and we interpret this phenomenon as the different
metric adopted in their respective definition.

As the MIN is a well-defined measure of nonlocal correlations, and
investigations of its properties in different quantum systems will
provide us with useful information which includes its controlling
methods, its potential applications and other aspects related to the
fundamental problem of quantum mechanics, we hope our work presented
here may shed light on understanding MIN for the quantum spin
system, and stimulate related investigations in this field.

\section*{ACKNOWLEDGMENTS}
This work was supported by Natural Science Foundation of China under
Grant Nos. 11174165 and 11275099.

\newcommand{\PRL}{Phys. Rev. Lett. }
\newcommand{\RMP}{Rev. Mod. Phys. }
\newcommand{\PRA}{Phys. Rev. A }
\newcommand{\PRB}{Phys. Rev. B }
\newcommand{\NJP}{New J. Phys. }
\newcommand{\JPA}{J. Phys. A }
\newcommand{\JPB}{J. Phys. B }
\newcommand{\PLA}{Phys. Lett. A }
\newcommand{\NP}{Nat. Phys. }
\newcommand{\NC}{Nat. Commun. }
%

%

\end{document}